\documentclass{article}
\usepackage{spconf,amsmath,graphicx}
\usepackage{multirow}
\usepackage{tabularx}
\usepackage{adjustbox}
\usepackage{graphicx,relsize,afterpage,textcomp,xcolor,soul,sfmath,authblk,mathtools,multicol,float,enumitem,titlesec,wrapfig,placeins}
\definecolor{dblue}{rgb}{0,0,1}

\usepackage[colorlinks=true,linkcolor=dblue,citecolor=dblue,urlcolor=dblue]{hyperref}
\newfloat{efigure}{htp}{efig}
\floatname{efigure}{Extended Data Fig.}

\usepackage[nameinlink]{cleveref}
\Crefname{equation}{Equation}{Equations}
\Crefname{figure}{Figure}{Figures.}
\Crefname{efigure}{Extended Data Figure}{Extended Data Figures.}
\Crefname{tabular}{Table}{Tables}
\crefrangelabelformat{equation}{#3#1#4-#5#2#6}
\creflabelformat{equation}{#2#1#3}

\usepackage{enumitem}
\setlist{nosep, leftmargin=14pt}

\usepackage{mwe}
\usepackage{xcolor}

\usepackage{subfig}
\title{Towards unraveling calibration biases in medical image analysis}
\name{María Agustina Ricci Lara$^{1,2}$ \qquad Candelaria Mosquera$^{1,2}$ \qquad Enzo Ferrante$^{3}$ \qquad Rodrigo Echeveste$^{3}$}
\address{\normalsize{$^{1}$ Health Informatics Department, Hospital Italiano de Buenos Aires, Buenos Aires, Argentina}\\
\normalsize{$^{2}$ Universidad Tecnológica Nacional, Buenos Aires, Argentina}\\
\normalsize{$^{3}$ Research Institute for Signals, Systems and Computational Intelligence sinc(i) (FICH-UNL/CONICET), Santa Fe, Argentina}}
\mail{\small Correspondence to: maria.ricci@hospitalitaliano.org.ar, \{recheveste,eferrante\}@sinc.unl.edu.ar}

\begin{document}
\maketitle

\begin{abstract}
In recent years the development of artificial intelligence (AI) systems for automated medical image analysis has gained enormous momentum. At the same time, a large body of work has shown that AI systems can systematically and unfairly discriminate against certain populations in various application scenarios. These two facts have motivated the emergence of algorithmic fairness studies in this field. Most research on healthcare algorithmic fairness to date has focused on the assessment of biases in terms of classical discrimination metrics such as AUC and accuracy. Potential biases in terms of model calibration, however, have only recently begun to be evaluated. This is especially important when working with clinical decision support systems, as predictive uncertainty is key for health professionals to optimally evaluate and combine multiple sources of information. In this work we study discrimination and calibration biases in models trained for automatic detection of malignant dermatological conditions from skin lesions images. Importantly, we show how several typically employed calibration metrics are systematically biased with respect to sample sizes, and how this can lead to erroneous fairness analysis if not taken into consideration. This is of particular relevance to fairness studies, where data imbalance results in drastic sample size differences between demographic sub-groups, which, if not taken into account, can act as confounders.
\end{abstract}
\begin{keywords}
Fairness, Bias, Calibration, Skin Lesion Analysis
\end{keywords}
\section{Introduction}
\label{sec:intro}
Different studies have shown that data-driven algorithms can perform differently across subpopulations, and even intensify historically observed biases to the detriment of particular groups \cite{buolamwini2018gender,zou2018ai, ricci2022addressing}. In this sense, biased algorithms are defined as those that perform unevenly when evaluated in sub-groups defined in terms of a protected attribute, such as sex, gender, age, ethnicity, socioeconomic factors, among others. This behaviour has raised an alarm in the scientific community and the evaluation of algorithmic fairness \cite{mehrabi2021survey} has been gaining increasing attention.

As artificial intelligence (AI) tools start to obtain approval from the Food and Drug Administration in the US and the CE  (French for "European conformity") mark in Europe, allowing their use in clinical practice and showing an impact in patient outcomes \cite{esteva2021deep}, the evaluation of fairness stands out as an essential step for trustworthy systems. In this context, algorithmic bias has been previously analyzed in the healthcare field \cite{chen2021ethical} and in medical imaging particularly \cite{ricci2022addressing,larrazabal2020gender}. For example, Burlina et al. \cite{burlina2021addressing} demonstrated that the accuracy of a baseline deep learning model trained with a major public dataset for the detection of diabetic retinopathy was worse for dark-skinned individuals than for light-skinned. Similarly, Seyyed et al. \cite{seyyed2021} showed how hispanic women can be systematically underdiagnosed by algorithms trained with well-known open databases for the automated analysis of chest X-rays. Furthermore, Larrazabal et al. \cite{larrazabal2020gender} investigated the impact of gender imbalance in X-ray databases used to train computer-assisted diagnosis systems and found that this can result in decreased model performance for underrepresented groups.

While the number of studies of fairness in the field of Medical Imaging Computing (MIC) has started to increase, a recent commentary \cite{ricci2022addressing} showed that there are still significant areas of vacancy to be explored. These include auditing models aimed at processing different imaging modalities to solve problems in other medical specialties, and addressing questions regarding the properties of certain evaluation metrics in the context of healthcare. Considering that in algorithmic fairness studies, in general, there is a majority group and an underrepresented minority group, if the metrics selected for the assessment are biased with respect to the sample size, it is possible that the results hide unfavorable behavior of AI models for the latter group.  Moreover, few cases of potential bias assessment in terms of calibration have been observed in this domain in particular \cite{ricci2022addressing}.

%and issues to be resolved. For example, there are certain metrics defined in the fairness literature (e.g. statistical parity) which do not directly apply when dealing with 

%En un contexto de fairness donde el desabalance en at protegido impacta en target, entonces metricas podrían ocultar

Classification performance should be understood as the combination of discrimination and calibration capabilities \cite{blattenberger1985separating}. A system is said to be well-calibrated when its output reflects the uncertainty about a sample's class given the input information \cite{ovadia2019can,dawid1982well}; for example, when the output can be interpreted as the probability that the sample belongs to the class of interest given the input \cite{naeini2015obtaining}. A distinctive characteristic of the medical imaging domain is the need for interpretable outputs to assist in clinical decision-making. In this regard, quantifying model uncertainty becomes crucial \cite{uncertaintyNature}. As with discrimination metrics, the selection of calibration metrics for auditing algorithmic fairness in medical imaging must also be made taking into account the large differences usually seen in the population sizes between the sub-groups that will be analyzed. It is important to examine the behavior of these metrics in such situations to ensure their effectiveness in detecting and mitigating biases that may disproportionately affect underrepresented groups. For instance, some authors have reported that one of the most widely used calibration estimators in the scientific community, known as Expected Calibration Error (ECE), is highly sensitive and heavily biased with respect to sample size, and becomes monotonically worse with fewer samples in the evaluation set \cite{gruber2022better}.

To our knowledge no studies have been carried out so far in the field of medical imaging analyzing the impact of sub-group underrepresentation in the outcome of algorithmic fairness audits in terms of both discrimination and calibration metrics. For this reason, we performed a fairness analysis with respect to calibration and discrimination metrics in the context of skin lesion classification. We studied a set of commonly used metrics analyzing their behavior in the context of a specific data imbalance problem which is light-skinned vs. dark-skinned cases in dermatological databases. The skin tone or phototype is usually expressed through a six-point scale known as the Fitzpatrick scale \cite{fitzpatrick1988validity}. Skin cancer prevalence is reported to be lower on dark-skinned individuals as a result of the photoprotection to ultraviolet radiation provided by the concentration and distribution of melanin in these subjects \cite{agbai2014skin}. Moreover, it is known that skin cancer is detected at a more advanced stage in this group, leading to a worse prognosis \cite{gupta2016skin}. These factors, combined with historical biases and insufficient representation of dark-skinned individuals in medical studies, may be some of the underlying causes for their frequent underrepresentation in the databases used to train machine learning models \cite{kleinberg2022racial}. Under this scenario, we set out to find out whether this imbalance in terms of protected attributes has an effect on the performance of the developed models.

With respect to AI models applied to the classification of skin lesions, fairness audits have previously assessed models on people with different skin tones. In this sense, Li et al. \cite{li2021estimating} found that vanilla AI models had a lower classification performance in dark-skinned individuals compared to light-skinned individuals. However, in other studies, no systematic pattern in the accuracy of the classifiers has been observed among people with different skin tone \cite{kinyanjui2020fairness, fitzpatrick17k}.

\textbf{Contributions:} In this work, we studied performance gaps both in terms of discrimination and calibration between light-skinned and dark-skinned individuals in the detection of malignant lesions in dermatological images. We performed the numerical experiments with an available open database of skin clinical images that include skin tone information of subjects \cite{padufes}. We observed that, although discrimination performance can show no significant difference between protected groups, calibration metrics naively used may appear to show that the model outputs do not represent uncertainty fairly between them. However, by matching sample sizes, we demonstrated that the differences observed from the naive estimators of calibration between sub-groups are actually spurious. In addition, we performed a series of synthetic experiments in which we emulated different de-calibration scenarios. Through these, we were able to replicate the observed behavior with the real data and to demonstrate that sample sizes play a key role in the outcome of comparative studies of calibration in settings with uneven sample sizes between sub-groups, such as typical fairness studies.
Moreover, we show that, as models are overall better calibrated, finite size effects become more noticeable, which is relevant in scenarios where models are post-hoc re-calibrated after training. Overall, our study sheds light on the importance of considering differences in sample sizes among population sub-groups when performing fairness studies, specially in cases where biased estimators are used to compute errors.

\section{Numerical experiments on real data}
\label{sec:RealDataExperiments}

In this work we address the problem of binary classification of benign and malignant skin lesions in clinical (i.e., not dermoscopic) images. We study algorithmic fairness in terms of discrimination and calibration considering the patient's phototype as the protected attribute.

\subsection{Data}
\label{ssec:RealData}

The analysis was carried out employing the open access database PAD-UFES-20 \cite{padufes}, originated in Brazil, which consists of 2,298 clinical images from 1,641 skin lesions belonging to 1,373 patients. Skin lesion are classified into six categories: three benign conditions (actinic keratosis, seborrheic keratosis and nevus) and three malignant conditions (basal cell carcinoma, squamous cell carcinoma and melanoma). Images come accompanied by metadata regarding lesions details and patient information, including skin phototype, which was collected during patients' medical appointments by a group of specialists \cite{padufes}. Only a subset of 1,494 images from the dataset include the skin tone on the Fitzpatrick scale as metadata. 

Regarding class balance with respect to the classification target (malignancy), PAD-UFES-20 includes 1,089 malignant lesions (47.38$\%$ of the cases). Regarding class balance with respect to the protected attribute, the  number of light-skinned subjects greatly exceeds the number of dark-skinned subjects in the dataset; there are approximately 20 times more light-skinned cases than dark-skinned cases. 

We randomly divided PAD-UFES-20 into training, validation and test subsets with a patient-level split. To stratify the split both by the classification target (benign or malignant) and the protected attribute (light, dark or unknown), we applied a two stratified K-Folds strategy of five iterations, totalling 25 runs. This ensures that the distribution of lesion malignancy shown in \Cref{fig:Malignancy_Distribution} and patient skin tone shown in \Cref{fig:SkinTone_Distribution} are approximately maintained in all subsets. As cases of unknown skin tone were later removed from the test set, the relative frequency of benign and malignant cases in the target variable in this subset was affected.

\subsection{Model training}
\label{ssec:ModelTraining}

We used a VGG-19 architecture \cite{simonyan2014very} pretrained on ImageNet, with Keras as framework. %The last layers were replaced by flatten, dense and dropout layers. Images were resized keeping the aspect ratio to fit the input size of the VGG and were normalized using the mean and standard deviation means of each channel in each of the datasets. To avoid overfitting, vertical and horizontal flips, random rotations and zoom were applied to the images during training. Regularization and dropout were applied as well. 
We performed image normalization with the channel's mean and standard deviation on each dataset, and applied horizontal flips, random rotations and zoom as data augmentation. 

We trained models using all skin tone cases combined, with weighted binary cross entropy as loss function and Adam as optimizer for 100 epochs. Validation loss was monitored with a patience of 20 epochs to avoid overfitting. We used a learning rate of 0.0005 and 32 images per batch.

\subsection{Platt scaling}
\label{ssec:PlattScaling}

We evaluated the effect of applying a monotonic transformation to the models outputs (i.e., not changing discrimination performance). We fitted Platt Scaling as calibrator (logistic regression) \cite{platt1999probabilistic} using the log likelihood ratios (LLRs) of the validation subset on each run. We trained the calibrators with all available samples in every validation set regardless of the value of the skin tone variable. The fitted parameters were then used to transform the scores of the corresponding test sets. The performance of these calibrated scores was compared between groups, and they were especially used to calculate the calibration loss, as shown in  \Cref{ssec:PeformanceEvaluation}.

\subsection{Performance evaluation}
\label{ssec:PeformanceEvaluation}

We computed classification performance metrics on the test set of each split. Metrics are presented separately for both protected groups, using boxplots showing the mean, median and interquartile range over the 25 runs. %\textcolor{red}{We report metrics before and after applying Platt scaling. — solo el claibration loss—} 

Regarding discrimination metrics, we report the area under the receiving operating characteristic curve $AUC_{ROC}$ 
% employed as it is one of the most commonly used metrics in fairness studies in medical imaging \cite{ricci2022addressing}. 
and the area under the precision-recall curve ($AUC_{PR}$). For comparison with prior works on skin imaging fairness \cite{kinyanjui2020fairness}, we also present the balanced accuracy computed using a threshold of 0.5. In addition, we report a modified version of the $AUC_{PR}$ that normalizes this metric considering the area under the curve for a random guess classifier, referred to as $AUC_{PRG}$ \cite{bugnon2019deep}. 

To assess the calibration performance solely, we fitted a linear logistic regression algorithm (Platt scaling) on every validation subset. The performance metrics of the transformed scores on each corresponding test set were used as reference to decompose proper scoring rules (PSR) into discrimination and calibration components \cite{ferrer2022analysis}: $PSR=PSR_{Platt}+\Delta PSR=PSR_{discrimination}+PSR_{calibration}$. We calculated the difference in cross-entropy between model outputs before calibration and their calibrated or Platt-transformed versions to obtain the calibration component (a.k.a. calibration loss) of the metric: $\Delta CE = CE-CE_{Platt}=CE_{calibration}$. We applied the same procedure with the Brier score: $\Delta Brier = Brier-Brier_{Platt}=Brier_{calibration}$.

We also calculated bin-based metrics that are usually reported in computer vision studies as calibration metrics: the expected calibration error (ECE) and the maximum calibration error (MCE) \cite{naeini2015obtaining}. In addition, we calculated the adaptive ECE (AdaECE) \cite{mukhoti2020calibrating}, which fits bin boundaries so that the number of samples is equal across bins.

We employed the Wilcoxon Signed Rank test for two related samples to evaluate significant differences (at a 0.05 level) between groups.

\subsection{Results}
\label{sec:ResultsReal}
\subsubsection{Discrimination and calibration performance across protected attributes on the original test set}
\begin{figure}[!ht]
\subfloat[\label{fig:Disc_Cal_A}]{\includegraphics[width=0.5\textwidth]{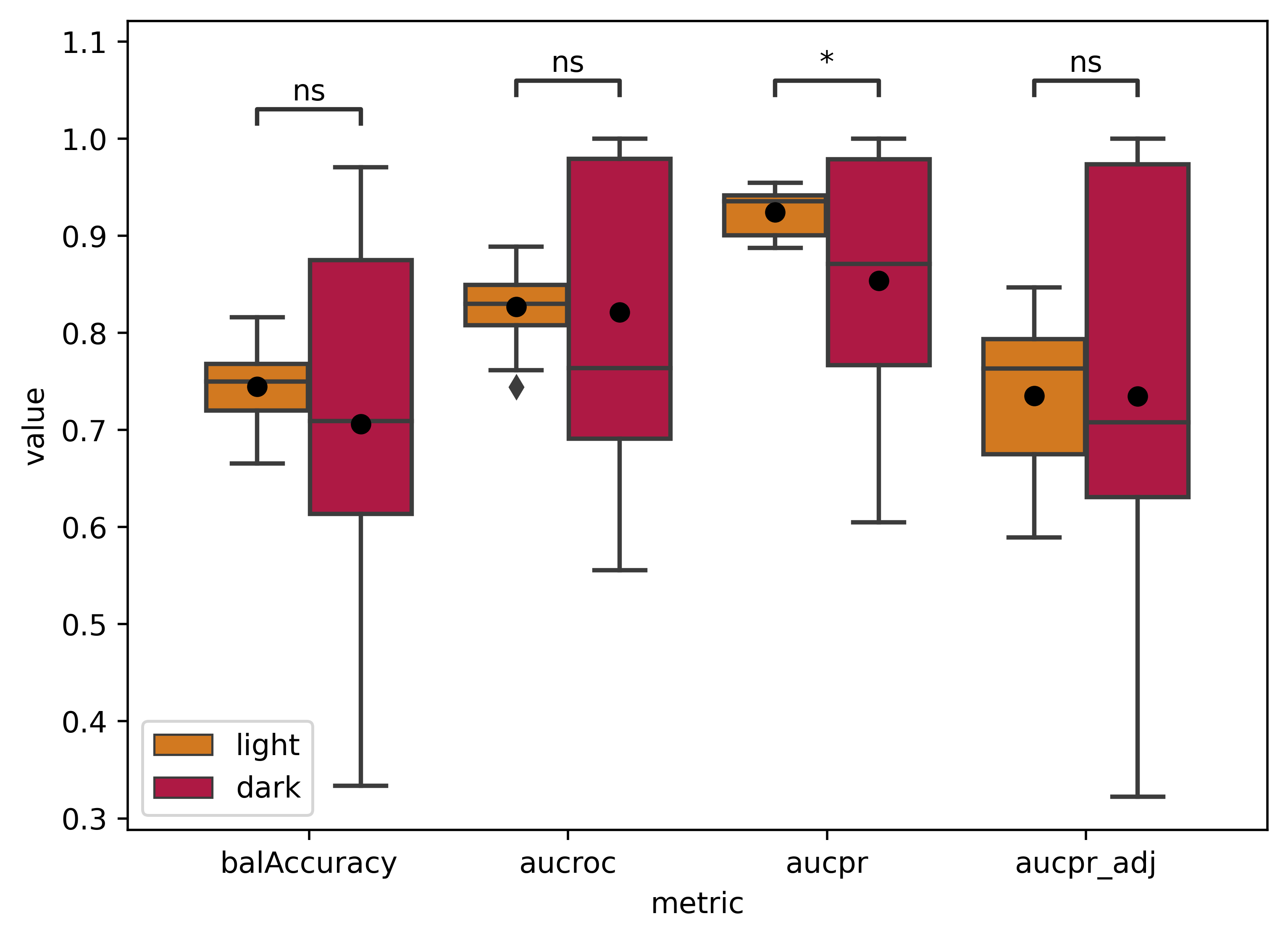}}\\
\subfloat[\label{fig:Disc_Cal_B}]{\includegraphics[width=0.5\textwidth]{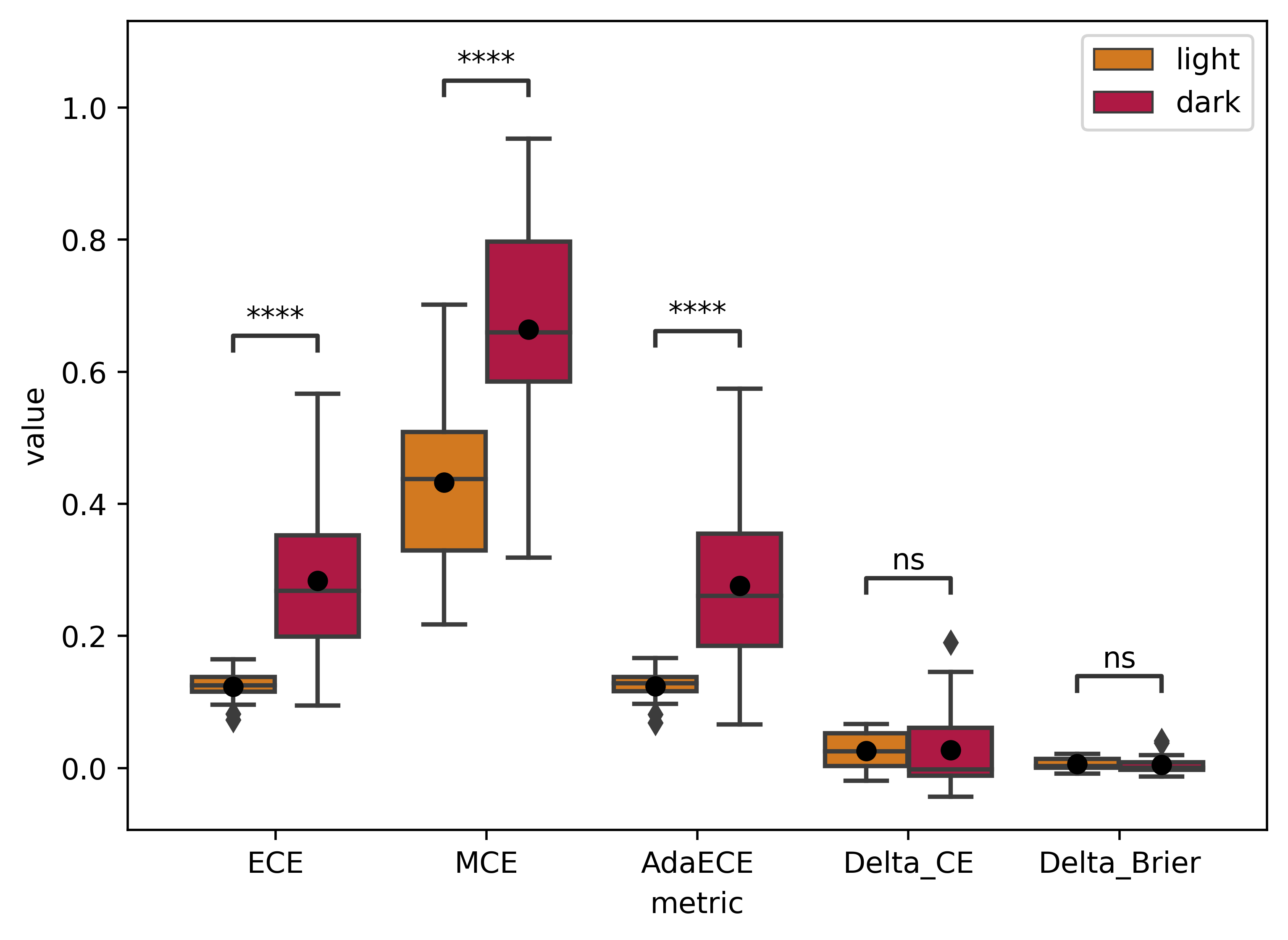}}
  \caption{\textbf{Comparison of metrics between sub-groups} Discrimination metrics \textbf{(A)} and calibration metrics \textbf{(B)} were computed over 25 runs. We note that calibration metrics were computed naively without taking sub-group sample sizes into account, resulting in \emph{spurious} significant differences}
    \label{fig:original_discrimination_calibration}
\end{figure}

Firstly we compared the performance between sub-groups with respect to discrimination metrics, which showed no statistical significance (\Cref{fig:Disc_Cal_A}). This is in line with previous studies such as \cite{kinyanjui2020fairness}. 

Next we compared sub-groups naively using calibration metrics without taking sample-sizes into account (\Cref{fig:Disc_Cal_B}), and found a trend favoring light-skinned individuals over dark-skinned individuals in many of them. Critically, as we will show in the next section, these differences were actually spurious, and a result of the different sample sizes between sub-groups.

\subsubsection{Sample size analysis on performance metrics}

\begin{figure}[!ht]
\subfloat[\label{fig:Cal_Min}]{\includegraphics[width=0.5\textwidth]{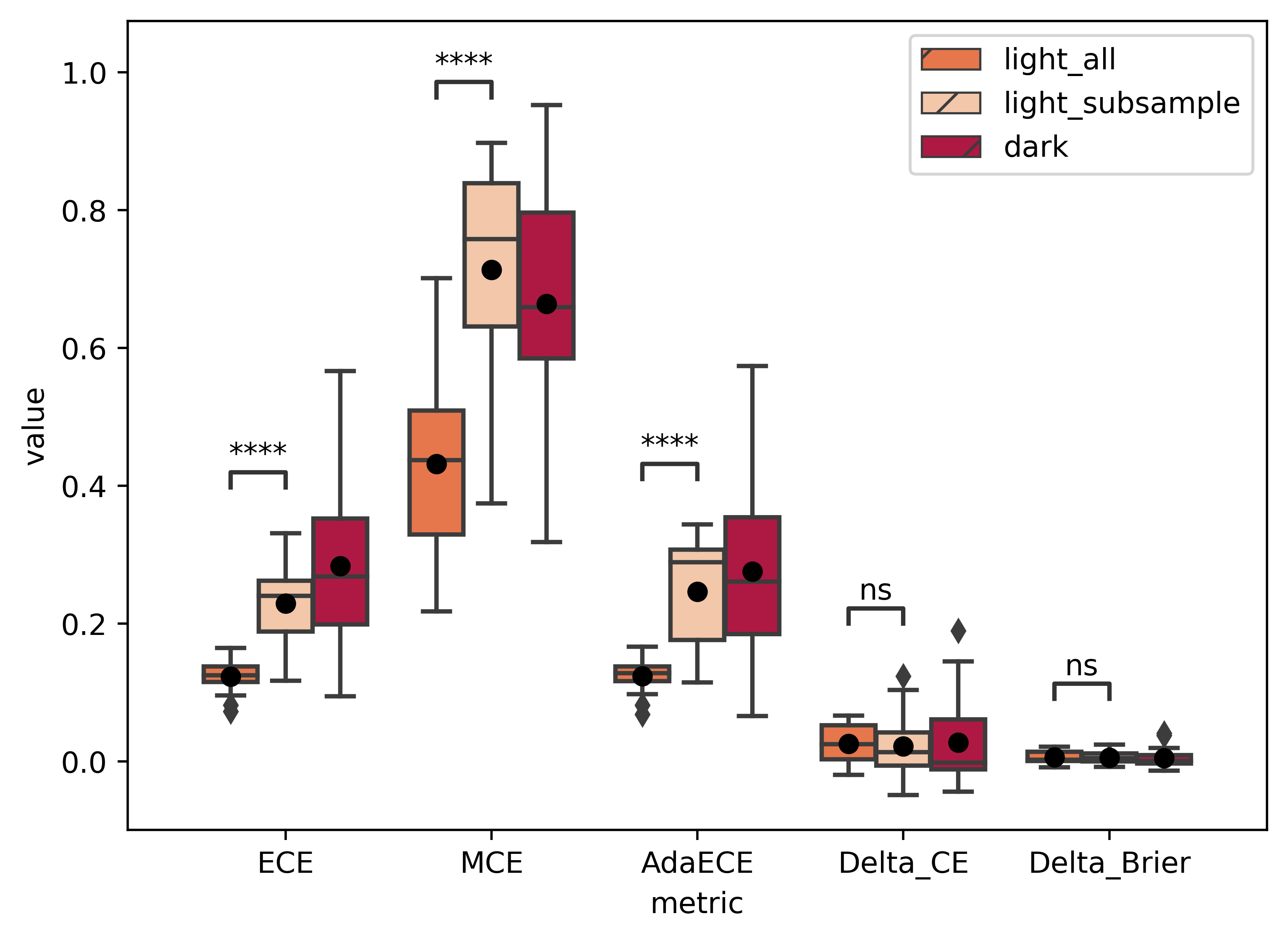}}\\
\subfloat[\label{fig:Cal_Equal}]{\includegraphics[width=0.5\textwidth]{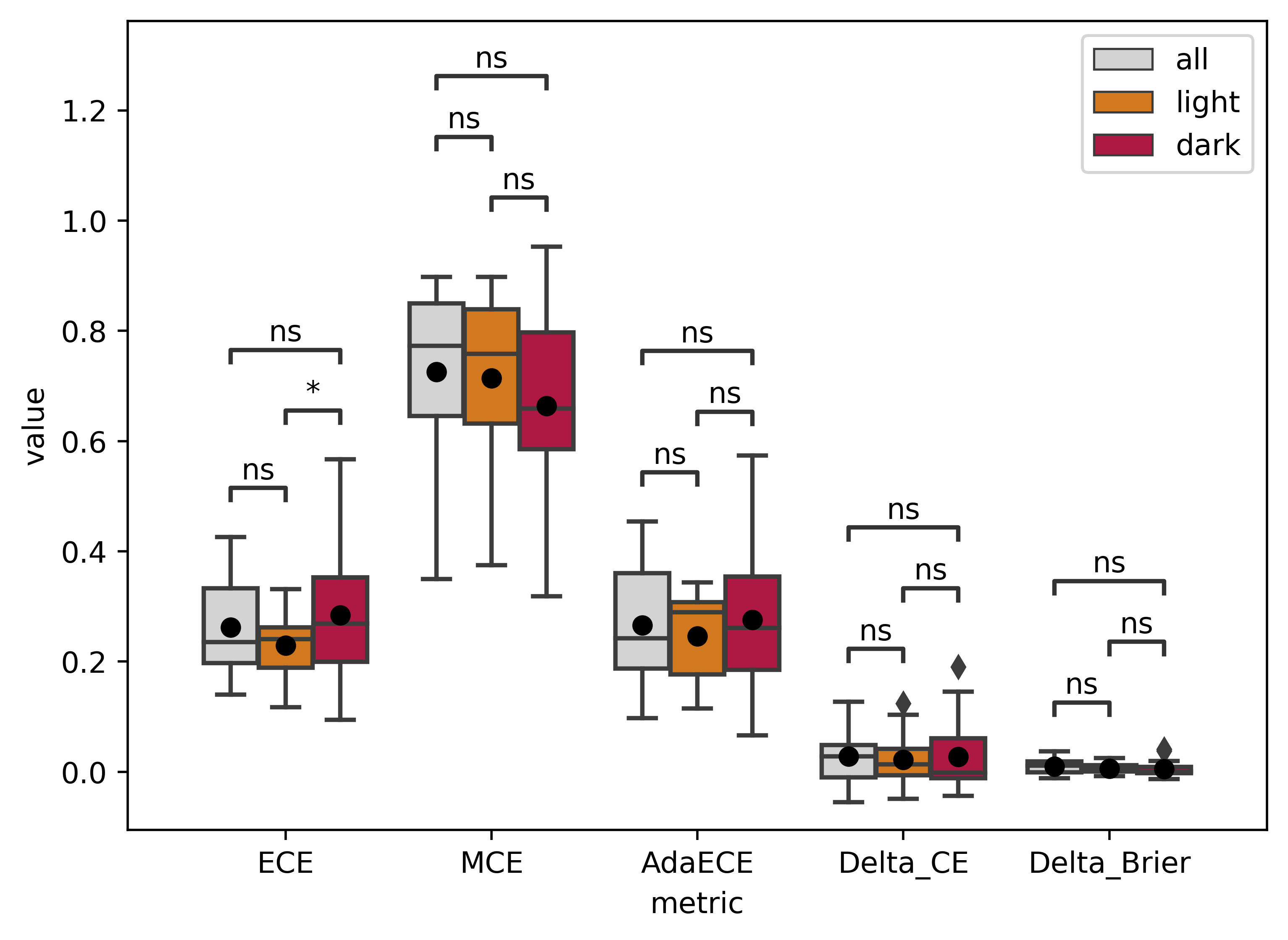}}
 \caption{\textbf{Effect of sample sizes on calibration metric fairness audits.} Significant differences on \textbf{(A)} show the impact of sample size in most of the metrics when comparing the original light-skinned test set and a sub-sampled version of it, matching the number of samples in the dark-skinned test set. When metrics are computed with equal sample sizes, they do not exhibit such differences \textbf{(B)}.}
    \label{fig:calibration_original_min_equal}
\end{figure}

As shown in \Cref{fig:dataset_distribution}, there are very few cases belonging to the minority class (dark-skinned individuals) in the test set. This situation, in fact, is manifested in the amplitude of the whiskers of the discrimination (\Cref{fig:Disc_Cal_A}) and calibration (\Cref{fig:Disc_Cal_B}) metric boxplots plotted for this sub-group in comparison with the light-skinned tone sub-group. We further analyzed whether this imbalance can influence the significant differences observed in the calibration metrics. By subsampling the larger group so that the number of samples matches that of the minority group, we find that previously observed differences are eliminated (\Cref{fig:calibration_original_min_equal}). %\Cref{fig:Cal_Min}, \Cref{fig:Cal_Equal}

\begin{figure*}
\centering
\subfloat[\label{fig:Boxplots_Skin}]{\includegraphics[width=1.0\textwidth]{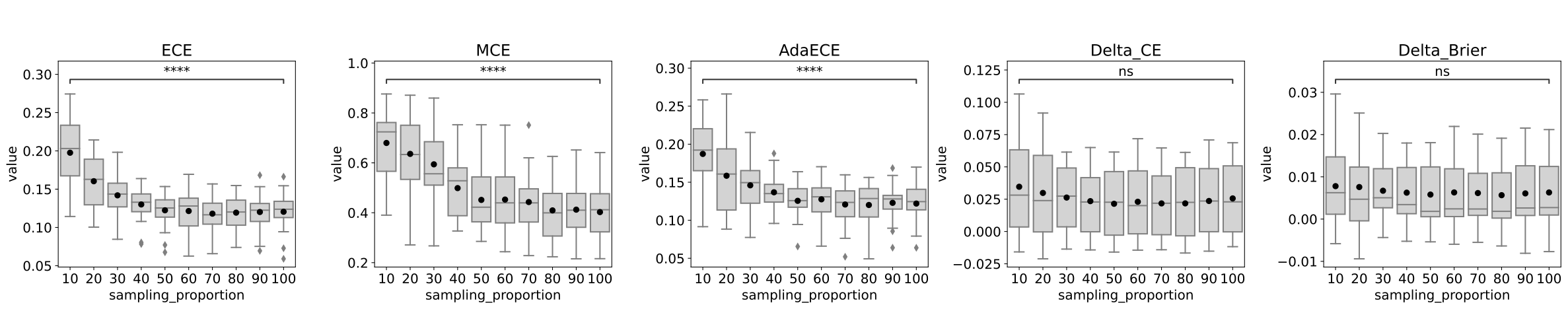}}\\
\subfloat[\label{fig:Boxplots_Synthetic}]{\includegraphics[width=1.0\textwidth]{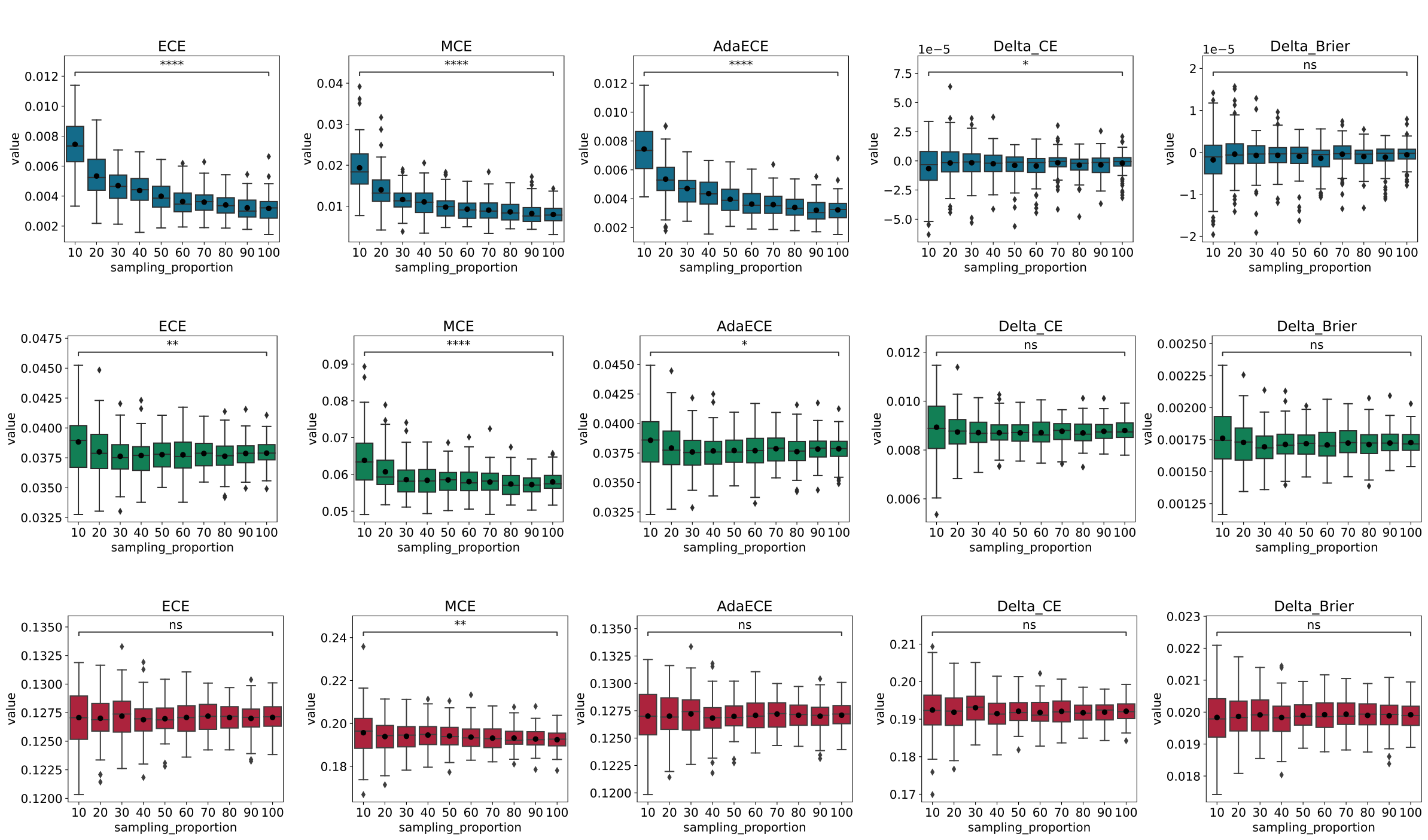}}
\caption{\textbf{Calibration metrics calculated over the 25 runs with different sampling ratios of the original test sets.} Results of the numerical experiments on real data \textbf{(A)} show the trend on ECE, MCE and AdaECE values with respect to the sampling proportion.  The results of the synthetic experiments \textbf{(B)} are arranged according to the different de-calibration scenarios: perfectly calibrated ($\alpha$ = $\beta$ =1) in the top row, slightly out of calibration ($\alpha$ = $\beta$ = 1.5) in the middle row and highly uncalibrated ($\alpha$ = $\beta$ = 5) in the bottom row.}
\label{fig:boxplots_sampling_proportion_calibration}
\end{figure*}

In addition, we show how randomly subsampling the entire data set with different sampling ratios affects the value of most of the calibration metrics, which, if not carefully taken into account, may give rise to erroneous or misleading findings (\Cref{fig:Boxplots_Skin}).

\section{Synthetic experiments}
\label{sec:SyntheticExperiments}

\subsection{Data}
\label{ssec:SyntheticData}

When subsampling, previously significant results may vanish either because a metric is biased with respect to sample size, or simply because the variances of the estimators increase and one has lost statistical power. The second component will necessarily always be present in part, but our hypothesis is that the first reason is playing a key role, causing spurious differences. In order to test this hypothesis we constructed a synthetic dataset, where these different factors can be independently controlled. 
We construct a binary classification problem where we know the true posteriors for each sample, and can manipulate calibration and sample size at will. To this end, we began with the creation of a set of one million samples that were randomly assigned to one of two categories (target variable). Random values associated with these samples were also generated to emulate the output scores of a model. Likewise, the cumulative distribution function of beta-distribution was used to provide different scenarios of model de-calibration by varying their $\alpha$ and $\beta$ parameters (\Cref{fig:betas}). At this point it should be noted that this transformation does not affect the value of the discrimination metrics like the balanced accuracy as the cut-off point is kept at 0.5. 

\subsection{Performance evaluation}
\label{ssec:SyntheticMethods}

The emulated scores in the dataset were modified by each of the beta functions describing the synthetic scenarios. For each of this cases, 100 random splits were performed to obtain train, validation and test sets. For every run, the validation set was used to fit a calibrator by Platt Scaling and the parameters obtained were used to calibrate the corresponding test set data. Later, the test set was sampled (with different sampling proportions, from 10\% to 100\%) to study the metrics behavior with respect to test size. Calibration metrics ECE, MCE and AdaECE were computed with the pre-calibration scores and calibration PSRs were obtained from the difference between the metrics (CE and Brier) computed with the pre- and post-calibration scores. 

The Wilcoxon Signed Rank test was used to test for significant differences (at a 0.05 level) between the values of the metrics calculated using the whole test size and the 10\% of it.

\subsection{Results}
\label{ssec:SyntheticResults}

The boxplots for the calibrated scenario (top row in \Cref{fig:Boxplots_Synthetic}), reveal a clear trend where ECE, MCE and AdaEce are biased with respect to the sampling ratio (i.e. set size), supported by the significant differences found between the value of these metrics computed on 100\% and 10\% of the data. Likewise, it can be observed that sample size does not seem to have a major effect on the PSRs. 

Moreover, when considering the rest of the scenarios (middle and bottom row in \Cref{fig:Boxplots_Synthetic}), it can be seen how the differences become less pronounced as de-calibration becomes more prominent. Comparing the scales on the y-axis we note that the finite size effect component in the metrics is eventually dominated by the decalibration component itself. 

\section{Discussion}
\label{sec:majhead}

Algorithmic fairness analysis can have important implications for defining the best way to present model results when this may impact clinical decision making, as pointed out in \cite{esteva2021deep}. Obtaining a fair calibration performance across demographic groups allows to use a common fixed threshold to obtain binary predictions that will imply the same error costs for all groups. On the contrary, if a common threshold is applied when the relationship between model outputs and uncertainty is different across protected groups, the majority group will likely have a real diagnostic performance that is closer to the expected one while the minority group will have unexpected false positive and negative rates.

Moreover, if in practice it were required to combine probabilistic sources of information (where the uncertainty of the combined prediction is dominated by the source of lowest uncertainty as stated in \cite{bejjanki2011cue}) and the models were uncalibrated or uncertainties were ignored when combining such sources, the error rate would be higher on average. Despite this important relationship between calibration and fairness in the context of healthcare, to our knowledge our work is the first to evaluate fairness in dermatological imaging AI using calibration metrics. 

Our experiments suggest that, although no differences in discrimination metrics were observed, differences in terms of calibration may appear when these metrics were used naively, without taking relative sample sizes into account. However, on closer inspection, taking into account the extreme imbalance in the data of the different sub-groups, as often is the case in fairness studies, we observe that these differences may actually be spurious, serving as a cautionary tale for researchers conducting these types of audits. Indeed it is possible to show how some of the metrics used in the literature present a high susceptibility to the size of the evaluation set. Likewise, it is possible to identify metrics that are less affected by this finite size effect and that could be more suitable to study biases in terms of calibration. On the other hand, we found that sample size dependence quickly ceases to be evident in the face of more strongly uncalibrated model scenarios. This suggests that the metrics could be interpreted by two components: a finite size effect component and another component that purely expresses model de-calibration. This raises a second warning, since models are these days usually not left uncalibrated, but re-calibrated, bringing them to the regime where calibration metrics become more sensitive to sample size.

As for experiments with real data, we face the usual limitation of having a small number of minority group samples available in open databases. When it comes to comparing sub-groups defined by different skin tones, there are few databases where this information is declared. In those few cases when phototype is reported, it is either the result of the evaluation of the image by specialists \cite{fitzpatrick17k} or a post-hoc estimation by computational algorithms such as the individual typological angle (ITA) \cite{kinyanjui2020fairness,li2021estimating}, which has recently been called into question \cite{groh2022towards}.  We were only able to use one database in which the skin tone label was determined following the usual protocol for patient assessment in the physician's office (PAD-UFES-20 \cite{padufes}), and this database has only 1,494 images for which the sensitive attribute of interest is reported. This dataset, moreover, has a significant imbalance between light-skinned and dark-skinned samples, intensifying the lack of images representing the minority group in the training, validation, and test partitions. This precluded performing counterfactual experiments to investigate whether extreme imbalance in skin tones during algorithm training affected algorithm performance when evaluated across all sub-groups. 

Overall, there is a clear need for algorithmic fairness studies in MIC that address calibration and not just discrimination performance. Our study raises a clear warning that these studies cannot be performed naively with any metric, without taking sample sizes into account. If sample sizes are different, but the minority group has enough samples to provide adequate measures of calibration, resampling may be a viable strategy. However, when samples are too scarce, robust metrics such as Delta-Brier or Delta-CE, may provide a more adequate solution. Moreover, we have shown that sample-size biases interact with overall model calibration. Biases become more marked as the model's overall calibration improves, which is relevant in scenarios where models are re-calibrated after training. Finally, our study provides yet another reason to emphasize the importance of inclusive datasets that enable comprehensive assessments of model fairness, including calibration, to ensure accurate and equitable machine learning outcomes.

%Another possible source of bias 
%Another situation occurs when an image classification problem is inherently more challenging for a sub-group than for others, as reported for chest x-ray analysis \cite{ganz2021assessing}. Regarding skin cancer, prevalence is usually higher in fair-skinned patients than in those with darker phototypes, as the concentration and distribution of melanin in the latter provides a photoprotective effect against damage caused by exposure to ultraviolet radiation. Moreover, in dark-skinned individuals the detection of skin cancer (and consequently, its inclusion in databases) is usually performed in advanced stages where the visual particularities of these lesions are more evident \cite{gupta2016skin, shue2022skin}. The extent to which these features of the task's setting influence algorithmic bias is yet to addressed.

\section{Acknowledgments}
\label{sec:acknowledgments}
We thank the Artificial Intelligence and Data Science in Health Program at Hospital Italiano de Buenos Aires for providing the space to discuss and work on these issues. This work was supported by Argentina's National Scientific and Technical Research Council (\mbox{CONICET}), which covered the salaries of R.E. and E.F. The work of E.F. was partially supported by the ARPH.AI project funded by a grant (Number 109584) from the International Development Research Centre (IDRC) and the Swedish International Development Cooperation Agency (SIDA). We also acknowledge the support of Universidad Nacional del Litoral (Grants CAID-PIC-50220140100084LI, 50620190100145LI), Agencia Nacional de Promoción de la Investigación, el Desarrollo Tecnológico y la Innovación (Grants PICT 2018-3907, PRH 2017-0003, PICT-2020-SERIEA-01765, PRH 2022-00002) and Santa Fe Agency for Science, Technology and Innovation (Award ID: IO-138-19). This work is supported by the Google Award for Inclusion Research (AIR) Program. We also thank Luciana Ferrer and Celia Cintas for the fruitful discussions. 

% References should be produced using the bibtex program from suitable
% BiBTeX files (here: strings, refs, manuals). The IEEEbib.bst bibliography
% style file from IEEE produces unsorted bibliography list.
% ------------------------------------------------------------------------- 
\bibliographystyle{IEEEbib}
\bibliography{refs.bib}

\newpage
\onecolumn 
\setcounter{figure}{0}
\renewcommand\thefigure{S\arabic{figure}} % modify \thefigure, *not* \figurename
\section*{Supplementary Figures}
\label{section:supplementary}

\begin{figure*}[ht]
\centering
\subfloat[\label{fig:Malignancy_Distribution}]{\includegraphics[width=0.5\textwidth]{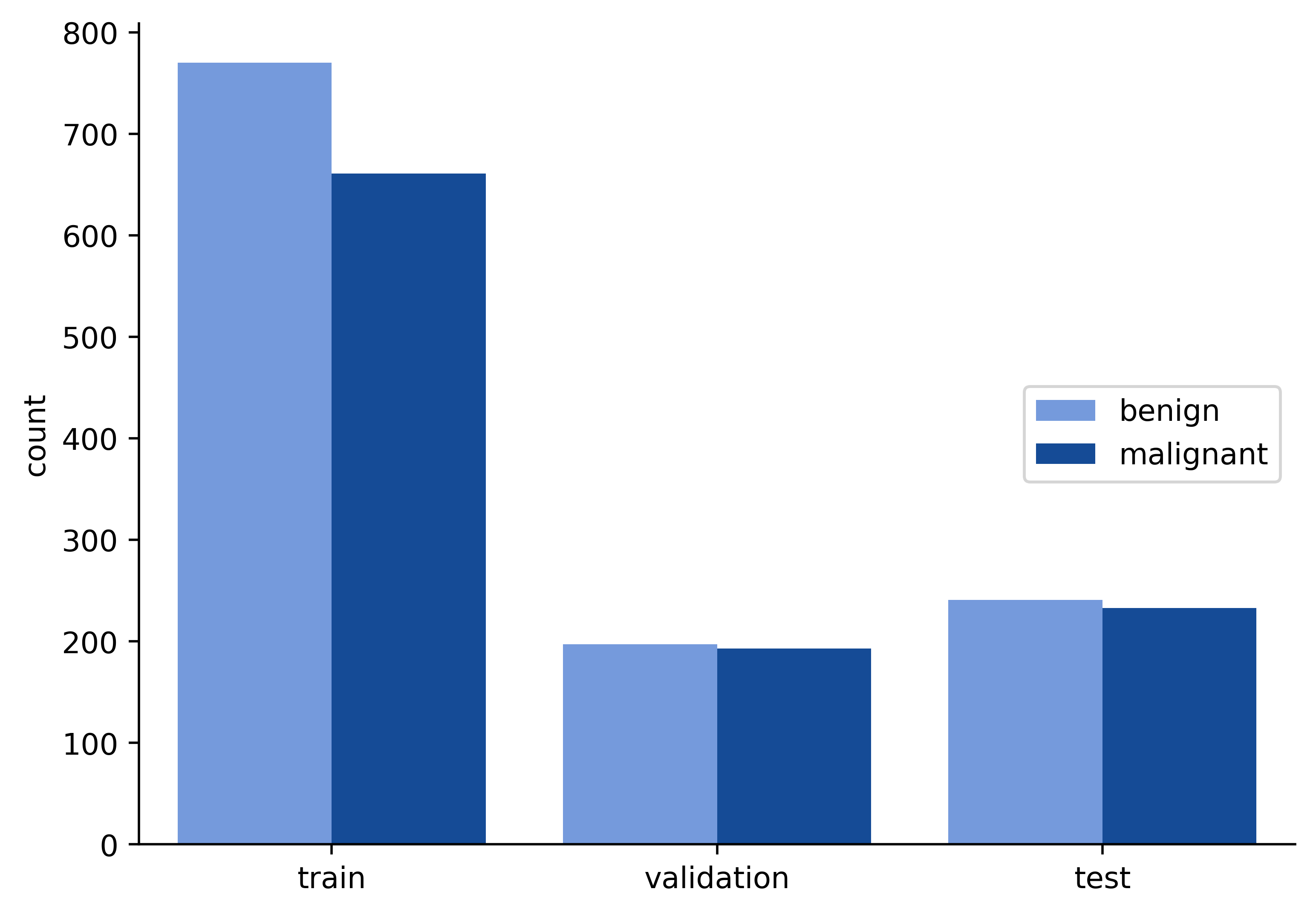}}
\subfloat[\label{fig:SkinTone_Distribution}]{\includegraphics[width=0.5\textwidth]{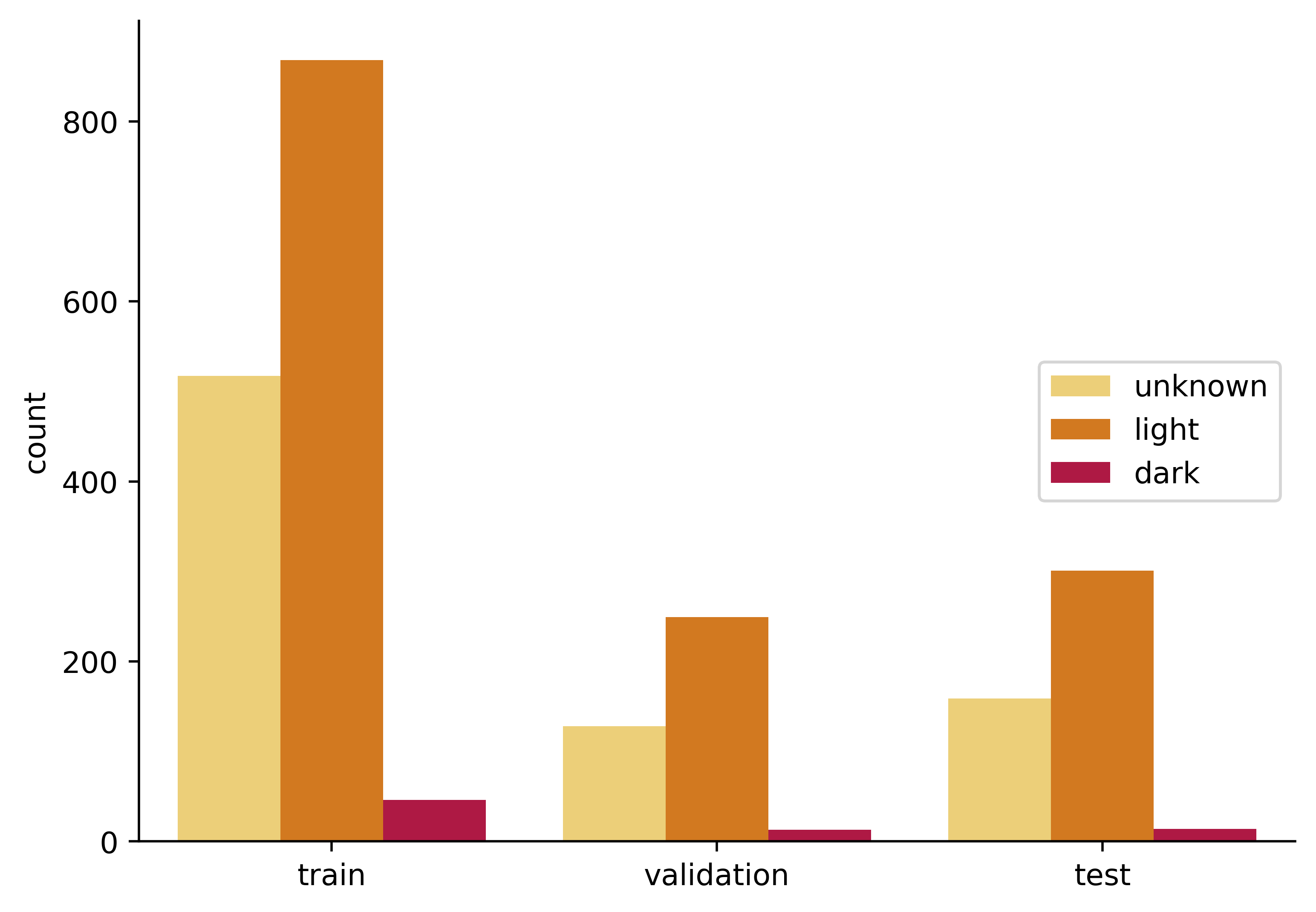}}
\caption{\textbf{Approximate distribution of samples in each subset.} Subsets compositions for outcome variable \textbf{(A)} or malignancy and protected attribute or skin tone \textbf{(B)}.}
\label{fig:dataset_distribution}
\end{figure*}

\begin{figure*}[ht]
\centering
\subfloat[\label{fig:Synthetic_betas}]{\includegraphics[height=4cm]{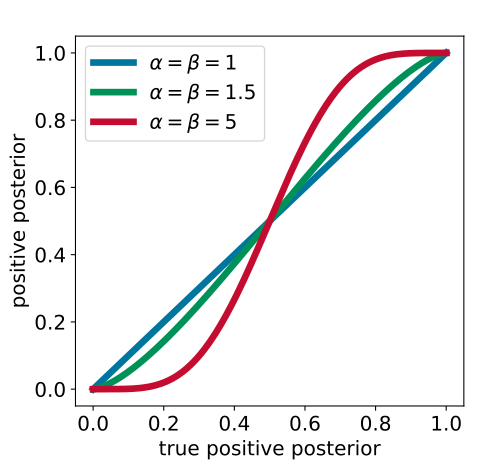}}
\subfloat[\label{fig:Synthetic_histograms}]{\includegraphics[height=4cm]{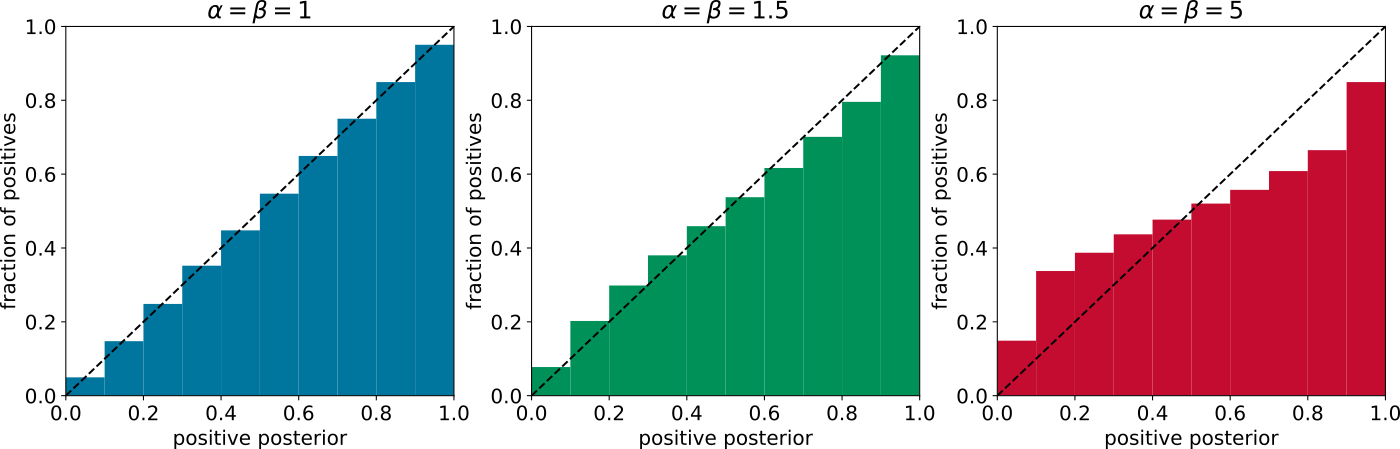}}
\caption{\textbf{Synthetic scenarios.} Cumulative distribution functions of beta-distribution \textbf{(A)} employed to create the calibrated ($\alpha$ = $\beta$ =1), slightly out of calibration ($\alpha$ = $\beta$ = 1.5) and highly uncalibrated ($\alpha$ = $\beta$ = 5) cases; and the corresponding reliability plots \textbf{(B)}.}
\label{fig:betas}
\end{figure*}

\end{document}